\documentclass[aps,amsfonts,nofootinbib,superscriptaddress]{revtex4}  

\usepackage{epsfig} 
\usepackage{feynmp}
\usepackage{appendix}
\usepackage{graphicx}
\usepackage{dsfont}
\usepackage{amssymb}
\usepackage{tipa}
\usepackage{amsmath}
\usepackage{amsbsy}
\usepackage{color}
\usepackage{slashed}
\usepackage{bm}
\usepackage{bbm}
\usepackage{float}
\usepackage{ulem}
\usepackage{tikz}
\usepackage{pgfplots}
\usepackage{xparse}
\usepackage{upgreek}
\usepackage{stmaryrd }
\usepackage{caption}
\usepackage{subfigure}
\usepackage{soul}
\usepackage{dutchcal}

\usetikzlibrary{positioning,arrows,patterns,automata}
\usetikzlibrary{decorations}
\usetikzlibrary{trees}
\usetikzlibrary{decorations.pathmorphing}
\usetikzlibrary{decorations.markings}
\usetikzlibrary{calc}

\newcommand{\ee}{\end{equation}}
\newcommand{\bb}{\begin{equation}}
\newcommand{\eqb}{\begin{eqnarray}}
\newcommand{\eqf}{\end{eqnarray}}

\tikzset{	aphoton/.style={decorate, decoration={snake}, draw=blue},
	photon/.style={decorate, decoration={snake}, draw=black},
	particle/.style={draw=black, postaction={decorate},
		decoration={markings,mark=at position .5 with {\arrow[draw=black]{>}}}},
	gluon/.style={decorate, draw=red,
		decoration={coil,amplitude=4pt, segment length=5pt}},
	vertex/.style={draw,shape=circle,fill=black,minimum size=3pt,inner sep=0pt},
}
\NewDocumentCommand\semiloop{O{black}mmmO{}O{above}}
{%
\draw[#1] let \p1 = ($(#3)-(#2)$) in (#3) arc (#4:({#4+180}):({0.5*veclen(\x1,\y1)})node[midway, #6] {#5};)}

\begin{document}
\title{Birefringence and hidden photons}
  \author{Ariel Arza}
\affiliation{Department of Physics, University of Florida, Gainesville FL 32611-8440, USA}
\email{ariel.arza@gmail.com}
\author{J. Gamboa }
\email{jorge.gamboa@usach.cl}
\affiliation{Departamento de  F\'{\i}sica, Universidad de  Santiago de
  Chile, Casilla 307, Santiago, Chile}

  \begin{abstract}
We study a model where photons interact with hidden photons and millicharged particles through a kinetic mixing term. Particularly, we focus in vacuum birefringence effects and we find a bound for the millicharged parameter assuming that hidden photons are a piece of the local dark matter density.
\end{abstract}
\date{\today}
\maketitle

\section{Introduction}

The vacuum  birefringence induced by external constant magnetic fields is a widely studied problem for several reasons \cite {vb1,vb2,vb3}, the first one is because the birefringence induced by external constant magnetic fields is experimentally interesting itself \cite{BFRT1} and the second one is because quantum electrodynamics in an external magnetic field is an ideal theoretical model where one could study axions \cite{maiani,vanbibber,raffelt,sikivie}. 

From the classical point of view the so called Cotton-Mouton effect is a phenomenon where a polarized light passes through a material in presence of a strong magnetic field and clear signal of birefringence appears \cite{rizzo1}. This  effect has been the starting point for many approaches and experiments trying to measure changes in polarization plane and ellipticity as signals of vacuum birefringence and existence of axions and millicharged particles \cite{PVLAS}.

Technically this problem is studied through the  effective (Euler-Heisenberg ) Lagrangean
\[
{\cal L}_{eff} = -\frac{1}{4} F^2 -i \ln ~\det \left(i \slashed D[A] -m_e\right),
\]
where the substitution $F_{\mu \nu} (A)\rightarrow  F_{\mu \nu} (A)+F^{ext}_{\mu \nu}$ is understood, with $F_{\mu \nu}(A)$ and 
$F^{ext}_{\mu \nu}$ the dynamical an external fields respectively. 
\medskip 

The explicit calculation yields to \cite {vb1,vb2,vb3}
\bb
{\cal L}_{eff}=-\frac{1}{4}F_{\mu\nu}(A)F^{\mu\nu}(A)+\frac{\alpha^2}{90m_e^2}\left[\left(F_{\mu\nu}(A)F^{\mu\nu} (A)\right)^2+\frac{7}{4}\left(F_{\mu\nu}(A)\tilde F^{\mu\nu} (A)\right)^2\right] + \cdots, \label{L1}
\ee
where  $\alpha =\frac{e^2}{4\pi}$ is  the fine structure constant. Thus, the presence of an external static magnetic field $B_0$ works like a birefringent medium with  parallel and perpendicular refractive $n_\parallel$ and $n_\perp$ indices and the difference between both is given by
\bb
\Delta n=n_\perp -n_{||}= \frac{\alpha}{30\pi}\left(\frac{B_0}{B_{cr}}\right)^2, \label{deltan1}
\ee
where $B_{cr}=m_e^2/e$.
\medskip

The study of modified Maxwell equations by axion matter has been performed in many papers \cite{maiani,vanbibber,raffelt,sikivie,nos} emphasizing different viewpoints  and, particularly, its  implications in optical experiments. In the seminal work \cite{raffelt} an exhaustive study was done and the consequences of axions observability discussed in connections with the ellipticity and rotation of the plane of polarization \cite{maiani}.
\medskip 

However one would like to go further and incorporate other fields such as the light sector of dark matter  by increasing the gauge group 
\bb
U(1) \rightarrow U(1)\times U'(1), \label{gauge1}
\ee
 where $U '(1)$ parameterizes the hidden photon sector \cite{holdom}. 
 \medskip 
 
 The gauge group (\ref{gauge1}) provides of direct way to incorporate interactions by using gauge invariance as basic criteria and, therefore by following the analogy with (\ref{L1}),  one should have an effective Lagrangean such as
 \bb 
 {\cal L} = {\cal L}_f -\frac{1}{4}F^2(A) -\frac{1}{4}F^2(B) + \frac{\xi_1}{2} F (A) F (B) + \frac{\xi_2}{2} \left(F(A) {\tilde F} (B)\right)^2 + \frac{\xi_3}{2} \left(F(A) F (B)\right)^2 + \cdots, \label{gauge2}
 \ee  
 where ${\cal L}_f$ the purely fermionic part, $\xi_{1,2,3}$ are unknown coefficients and $F(A)F(B) \equiv F_{\mu \nu} (A) F^{\mu \nu} (B)$ is the kinetic mixing term \cite{holdom}. 
 \medskip
 
 This effective Lagrangian leads to a set of modified Maxwell equations whose solutions are consistent with solutions in a refractive medium.
  
The purpose of this research will be study the effects due to birefringence and the possibility of observing them considering that the fields produced in the Maxwell equations are due to the presence of dark matter hidden photons sources as in the circuit-LC studied in \cite{LC,pao}\footnote{Although the literature in this field is very extense, see for example \cite{sikivie}.}.

\section{Hidden Sector Photons and Millicharged Particles} 

In order to carry out the idea outlined above let's starting  considering the following Lagrangean
\bb
{\cal L} ={\cal L}_{QED}  + {\cal L}_{hQED} + {\cal L}_{KM}, \label{total}
\ee 
where 
\eqb 
{\cal L}_{QED} &=& 
 \bar\psi(i\slashed\partial-e\slashed A-m_e)\psi -\frac{1}{4}F_{\mu\nu}F^{\mu\nu}, \label{qed1}
 \\
 {\cal L}_{hQED} &=& \bar\chi(i\slashed\partial-e_h\slashed A'-m_h)\chi 
-\frac{1}{4}F'_{\mu\nu}F'^{\mu\nu}+\frac{m^2}{2}A'_\mu A'^\mu, \label{dqed} 
\\
{\cal L}_{KM} &=&\frac{\xi}{2}F_{\mu\nu}F'^{\mu\nu}, \label{L00}
\eqf 
where $\chi$ and $A'_\mu$ are the hidden fermions and photons fields, respectively, and $m$ the hidden photon mass.
\medskip 

Integrating-out the fermions $\chi$ we obtain the following effective Lagrangean 
\eqb
{\cal L}_{eff} &=& {\cal L}_{QED}-\frac{1}{4}F'_{\mu\nu}F'^{\mu\nu}+\frac{m^2}{2}A'_\mu A'^\mu \nonumber
\\
& &+\frac{\xi}{2}F_{\mu\nu}F'^{\mu\nu}+\frac{\kappa}{8}\left[\left(F'_{\mu\nu}F'^{\mu\nu}\right)^2+\frac{7}{4}(F'_{\mu\nu}\tilde F'^{\mu\nu})^2\right] + \cdots, \label{L3}
\eqf
where
\bb
\kappa=\frac{1}{180\pi^2}\left(\frac{e_h}{m_h}\right)^4. \label{kappa}
\ee

However, the dynamical effects of the Lagrangian (\ref{L3}) in an external hidden electromagnetic field $F'_{0\mu\nu}$ are figured out  from the solutions of the equations of motion, namely

\eqb
&&\partial_\mu F^{\mu\nu} = -\xi\partial_\mu F'^{\mu\nu}+j^\nu, \label{eqA}
\\
&&\partial_\mu F'^{\mu\nu}+m^2A'^\nu = -\xi\partial_\mu F^{\mu\nu} +\kappa\partial_\mu H^{\mu\nu}, \label{eqA'}
\eqf
where $H^{\mu\nu}$ has been defined as 
\bb
H^{\mu\nu}=2F'_{0\alpha\beta}F'^{\alpha\beta}F_0'^{\mu\nu}+F'_{0\alpha\beta}F_0'^{\alpha\beta}F'^{\mu\nu}+\frac{7}{2}F'_{0\alpha\beta}\tilde F'^{\alpha\beta}\tilde F_0'^{\mu\nu}+\frac{7}{4}F'_{0\alpha\beta}\tilde F_0'^{\alpha\beta}\tilde F'^{\mu\nu}. \label{H1}
\ee

These are the full set of equations of motion once the hidden fermions have been integrated-out. The next step is more technical because we must solve the equations of motion by assuming that hidden photons make up the local dark matter. The birefringence effects in external electromagnetic fields have been studied, for example in \cite{ahlers}, but in this paper in the appendix, we will provide additional arguments in this directio. 

\section{Birefringence in the presence of Dark Matter as Hidden Photons}  \label{birefringence}

In order to find the effects of birefringence, we work on the photon-hidden photon oscillation picture where 
\[
a^\mu = {A'}^\mu + \xi A^\mu. 
\]
The equations (\ref{eqA}) and (\ref{eqA'}) transform to
\eqb
\partial_\mu F^{\mu\nu}+\xi^2m^2A^\nu &=& \xi m^2a^\nu-\xi\kappa\partial_\mu H^{\mu\nu}(a,A), \label{eqA1}
\\
\partial_\mu f^{\mu\nu}+m^2a^\nu &=& \xi mA^\nu+\kappa\partial_\mu H^{\mu\nu}(a,A), \label{eqA'1}
\eqf
where $f_{\mu\nu}=F'_{\mu\nu}+\xi F_{\mu\nu}$. Using the fact that $f^{\mu\nu}$ and $H^{\mu\nu}$ are antisymmetric tensors, the four-divergence of  (\ref{eqA'1})  implies $\partial_\mu a^{\mu}=\xi\partial_\mu A^{\mu}=0$ in the Lorentz's gauge, then the equations (\ref{eqA1}) and (\ref{eqA'1}) become
\eqb
(\Box+\xi^2m^2)A^\nu &=& \xi m^2a^\nu-\xi\kappa\partial_\mu H^{\mu\nu}(a-\xi A), \label{eqA2}
\\
(\Box+m^2)a^\nu &=& \xi m^2A^\nu+\kappa\partial_\mu H^{\mu\nu}(a-\xi A). \label{eqA'2}
\eqf 

One way to see birefringence effects is through the propagation of a laser beam. This could acquire a visible ellipticity and rotation in the polarization vector. If we initially have a linearly polarized laser spreading in the direction $\hat x$, the problem can be treated in one spatial dimension. Equations (\ref{eqA1}) and (\ref{eqA'1}) can be solved by perturbation theory, by defining   
\eqb 
A_\mu &=& A_\mu^{(0)}+\xi^2 A_\mu^{(2)}+\xi^4A_\mu^{(4)}+ \cdots \nonumber
\\
a_\mu&=& \xi a_\mu^{(1)}+\xi^3 a_\mu^{(3)}+\xi^5a_\mu^{(5)}+ \cdots, \label{perturb}
\eqf
where we have denoted the scalar and vector potentials as 
\[
a_\mu^{(i)}=(\varphi^{(i)},-\bm a^{(i)}), ~~~~~~~~~A_\mu^{(j)}=(\phi^{(j)},-\bm A^{(j)}), ~~~~i=1,3,\cdots, \ \ \ \ j=2,4,\cdots,
\]

 and $A_\mu^{(0)}$ is the free laser electromagnetic field defined as 
\bb
\bm A^{(0)}(x,t)=\bm\alpha e^{i\omega(x-t)}, \ \ \ \ \ \ \ \ \ \ \phi^{(0)}=0, \label{initial}
\ee
where $\bm\alpha$ is the initial polarization vector in the $yz$ plane and $\omega$ the frequency of the laser field. At this point, except for $A_\mu^{(0)}$, the fields should be found solving equations (\ref{eqA2}) and (\ref{eqA'2}) with the boundary conditions 
\bb
a_\mu^{(i)}(0,t)=A_\mu^{(j)}(0,t)=0. \label{bc1}
\ee

Hidden Photons has been proposed as a dark matter candidates \cite{arias1}. In such case, the dark matter field is described by a hidden electric field $D(t)$, which oscillates periodically at a frequency equal to the hidden photon mass and which is related to the local dark matter density $\rho_\text{DM}$ through $\left<D^2(t)\right>=2\rho_\text{DM}$, where $\left<\,\,\right>$ denotes a temporal average. We write the external hidden electromagnetic tensor and its dual as
\eqb
F'_{0\alpha \beta} &=& \left[ {\begin{array}{cccc}0&0&0&D(t) \\
0 &0&0&0\\
0&0&0&0 
\\ 
-D(t)&0&0&0\\ \end{array}}\right]\ , ~~~
{\tilde F}'_{0\alpha \beta} = \left[ {\begin{array}{cccc}0&0&0&0\\
0&0&D(t)&0\\
0&-D(t)&0&0 
\\ 
0&0&0&0\\ \end{array}}\right] \, .
 \eqf

Note that quantum effects will bee seen at second order in $\xi$ for $A^\nu$. At first order in $\xi$, (\ref{eqA'2}) becomes 
\bb
(\partial_t^2-\partial_x^2+m^2)\bm a^{(1)}=m^2\left(\bm\alpha -\frac{4\kappa D^2\omega^2}{m^2}\bm\beta\right)e^{i\omega(x-t)}, \label{eqad11}
\ee
where we have assumed that $D(t)$ is constant compared with the high frequency of the laser and $\bm{\beta}=\hat y\frac{7}{4}\alpha_y+\hat z\alpha_z$. Since the scalar potential does not receive extra contributions because there are no sources, we find $\varphi^{(1)}=0$. For the vector potential it is convenient to write ${\bf a}^{(1)} (x,t)= e^{-i \omega t} {\bf a}^{(1)} (x)$ by assuming that $m\ll \omega$ and to make the approximation $\partial_x^2 + \omega^2 = (-i \partial_x + \omega)(i\partial_x +\omega) \approx 2 \omega (i\partial_x + \omega)$. 

These assumptions transform (\ref{eqad11}) into   
\bb
(i\partial_x+\omega-q)\bm a^{(1)}(x)=-q\left(\bm\alpha-\frac{4\kappa D^2\omega^2}{m^2}\bm{\beta}\right)e^{i\omega x}, \label{eqad12}
\ee
where we have defined $q=\frac{m^2}{2 \omega}$.

The above equation is solved with the boundary condition (\ref{bc1}), we obtain 
\bb
\bm a^{(1)}(x)=\left(\bm\alpha-\frac{4\kappa D^2\omega^2}{m^2}\bm{\beta}\right)(1-e^{-iqx})e^{i\omega x}. \label{solad11}
\ee

\begin{figure}[t!]
\includegraphics[scale=0.8]{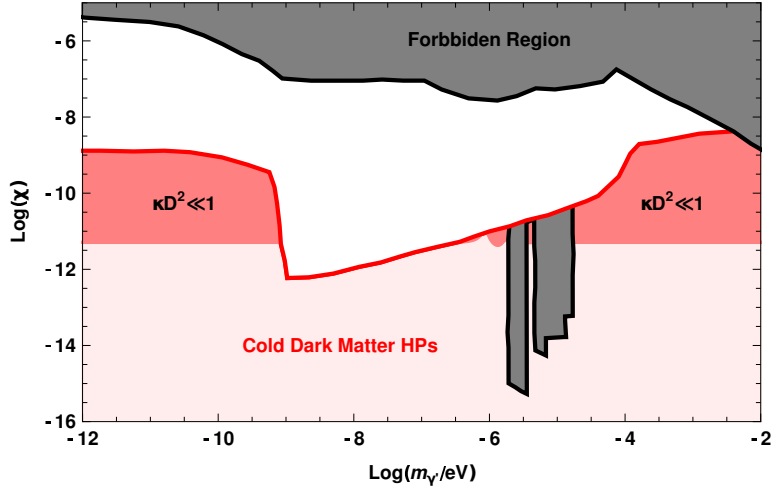}
  \caption{{\small This plot shows the regions of the space of parameters presently used for searching for hidden photons \cite{space}. The gray region corresponds to the discarded hidden photons one. The pink regions are the parameters space where these particles can be dark matter. The dark pink one is where our results are valid.}}
  \label{fig1}
\end{figure}

Following the same arguments above, at second order in $\xi$ we have $\phi^{(2)}=0$ and the spatial part of equation of motion (\ref{eqA2}) becomes 
\bb
(i\partial_x+\omega)\bm A^{(2)}(x)=q\left(\bm\alpha e^{-iqx}+\frac{4\kappa D^2\omega^2}{m^2}\bm{\beta}(1-2e^{-iqx})\right)e^{i\omega x} \label{eqAd11}
\ee 
whose solution is given by
\bb
\bm A^{(2)}(x)=-\left(\bm\alpha(1-e^{-iqx})-\frac{4\kappa D^2\omega^2}{m^2}\bm{\beta}(2-iqx-2e^{-iqx})\right)e^{i\omega x}. \label{solAd21}
\ee
When the polarization of a laser beam is described by the components $A_y=\alpha_y(1+\eta_y+i\sigma_y)e^{i\omega(x-t)}$ and $A_z=\alpha_z(1+\eta_z+i\sigma_z)e^{i\omega(x-t)}$, where $\sigma_{y,z}\ll1$ and $\eta_{y,z}\ll1$, the ellipticity $\varepsilon$ and rotation $\delta\theta$ are given respectively by
\bb
\varepsilon=\frac{\sin2\theta_0}{2}(\sigma_z-\sigma_y) \label{ellip1}
\ee
and
\bb
\delta\theta=\frac{\sin2\theta_0}{2}(\eta_z-\eta_y), \label{rot1}
\ee
where $\theta_0$ is the initial polarization angle. Taking the result (\ref{solAd21}) and assuming $\theta_0=\pi/4$ , we find an ellipticity
\bb
\varepsilon=\frac{3\xi^2\kappa\omega^2\rho_\text{DM}}{m^2}|qx-2\sin qx| \label{ellip3}
\ee
and a rotation
\bb
\delta\theta=\frac{6\xi^2\kappa\omega^2\rho_\text{DM}}{m^2}(1-\cos qx). \label{rot3}
\ee
Taking into account the last result of the PVLAS experiment \cite{PVLAS}, we can obtain a bound for $\epsilon/m_h$, where $\epsilon$ is the millicharge parameter defined in this case as $e_h\xi/e$. The experiment worked with laser frequency $\omega\sim1\text{eV}$ and used a Fabry-Perot cavity, which the effective path of the laser beam is $L=1.3\times10^6\text{m}$. It found no signals with a sensitivity of $\varepsilon\sim 10^{-10}$. On the other hand, the estimates of the local dark matter density is given by $\rho_\text{DM}=300\text{MeV}/\text{cm}^3$. With these data imply that 
\bb
\epsilon<9.82\times10^{-4}\sqrt{\xi}\left(\frac{m_h}{\text{eV}}\right). \label{bound}
\ee
This result is a bound for hidden sector fermions if dark matter is composed by hidden photons. It is important to mention that the bound (\ref{bound}) is valid only in a certain space of parameters $(\xi,m)$. Such a limitation is due to the fact that the calculations were made with the equations of motion provided by lagrangian (\ref{L3}), which was truncated at first order in $\kappa$. Thus, the perturbation theory remain valid in the limit $\kappa D^2\ll1$. This result and the bound (\ref{bound}) are consistent with  the parameters space showed in figure (\ref{fig1}) \cite{arias1,space}. Another concern that must be taking into account is that we have assumed a laser propagating orthogonally to the dark matter hidden electric field. We do not know whether the dark matter vector has a preferred direction in space or whether it is randomly oriented. In the two cases, we must to add a factor $\kappa=\sin^2\theta$ to the results in ellipticity and rotation, where $\theta$ is the angle between the laser beam and the dark matter electric field. As it was argued in reference \cite{pao}, if the hidden electric field has a preferred direction, we can take the conservative choice $\kappa=0.05$, where its real value is bigger with a 95\% confidence level. On the other hand, if this vector is randomly oriented, we average over all possible angles, thus $\kappa=0.5$.

\medskip

{\it Acknowledgements}:  This work was supported by USA-1555.
\appendix

\section{Birefringence in the presence of an external static electric field} \label{electric}

Let's consider an static electric field $\bm E=\hat zE_0$ inside two parallel conducting plates, separated by a distance $d$. If the electric field is normal to the plates plane, the total charge density in the plates is given by
\eqb
\rho &=& \bm\nabla\cdot\bm E=\hat z\partial_z(E_0\Theta(z)\Theta(d-z)) \nonumber
\\
&=& E_0(\delta(z)-\delta(z-d)). \nonumber
\eqf
Replacing (\ref{eqA}) into (\ref{eqA'}), one can see that charge density should induce a hidden electric field $\bm E'_0$ (as we will see later). At first order in $\xi$, the static equations associated to (\ref{eqA'}) is
\bb
(\partial_z^2-m^2)\phi'(z)=\xi E_0[\delta(z)-\delta(z-d)], \label{eqE0hidden}
\ee
whose solution is 
\bb
\bm \phi'(z)=\xi E_0\int_{-\infty}^\infty dz'G(z-z')[\delta(z')-\delta(z'-d)], \label{solB0hidden1}
\ee
where $G(z-z')$ is the Green's function of the problem given by $G(z-z')=-e^{-m|z-z'|}/(2m)$. For $0< z<d$, we have the solution $\phi'(z)=-\frac{\xi E_0}{2m}[e^{-mz}-e^{-m(d-z)}]$ and the hidden magnetic field is given by
\bb
\bm E_0'=-\hat z\partial_z\phi'(z)=-\hat z\frac{\xi E_0}{2}[e^{-mz}+e^{-m(d-z)}]. \label{solE0hidden}
\ee
If $md\ll 1$ the external hidden magnetic field is $-\hat z\xi E_0$ and, therefore, the electromagnetic tensor given by
\eqb
F'_{0\alpha \beta} &=& \left[ {\begin{array}{cccc}0&0&0&-\xi E_0 \\
0 &0&0&0\\
0&0&0&0 
\\ 
\xi E_0&0&0&0\\ \end{array}}\right]\, , ~~~
{\tilde F}'_{0\alpha \beta} = \left[ {\begin{array}{cccc}0&0&0&0\\
0&0&-\xi E_0&0\\
0&\xi E_0&0&0 
\\ 
0&0&0&0\\ \end{array}}\right] \, .
 \eqf  
 Proceeding as in section \ref{birefringence}, we find that vacuum birefringence effects occur at fourth order in $\xi$. The fourth order electromagnetic field is given by
\bb
A^{(4)}(x,t)=\bm\alpha[1-(1-iqx)e^{-iqx}]e^{i\omega(x-t)}+2i\kappa E_0^2\bm\beta\omega xe^{-iqx}e^{i\omega(x-t)}. \label{solAE4}
\ee
This leads to an ellipticity and rotation given by
\bb
|\varepsilon|=\frac{3}{4}\xi^4\kappa E_0^2\omega x|\cos(qx)| \label{ellip2}
\ee
and
\bb
|\delta\theta|=\frac{3}{4}\xi^4\kappa E_0^2\omega x|\sin(qx)|, \label{rot2}
\ee
respectively.

\section{Birefringence in the presence of an external static magnetic field}

Let's consider an static constant magnetic field in a region $0<x<L$ and pointing in the direction $\hat z$ then, the current density induced by this magnetic field is given by 
\eqb
\bm J&=&\bm\nabla\times\bm B=-\hat y\partial_x(B_0\Theta(x)\Theta(L-x)) \nonumber
\\
&=&-\hat yB_0(\delta(x)-\delta(x-L)). \nonumber
\eqf
Replacing (\ref{eqA}) into (\ref{eqA'}), one can see that current density should induce a dark magnetic field $\bm B'_0$. 

At first order in $\xi$, the static equations associated to (\ref{eqA'}) is 
\bb
(\partial_x^2-m^2)\bm A'=-\hat y\xi B_0[\delta(x)-\delta(x-L)]. \label{eqB0hidden}
\ee
The solution is found with the green's method such as in appendix \ref{electric}. For $0< x<L$, we have $\bm A'(x)=\hat y\frac{\xi B_0}{2m}[e^{-mx}-e^{-m(L-x)}]$ and the hidden magnetic field is 
\bb
\bm B'_0=\hat z\partial_xA'=-\hat z\frac{\xi B_0}{2}[e^{-mx}+e^{-m(L-x)}]. \label{solB0hidden2}
\ee
Therefore, we write the $F'$ and ${\tilde F}'$ tensors as 
\eqb
F'_{0\alpha \beta} &=& \left[ {\begin{array}{cccc}0&0&0&0 \\
0 &0& \frac{\xi B_0}{2} &0\\
0& -\frac{\xi B_0}{2}&0&0 
\\ 
0&0&0&0\\ \end{array}}\right] \, (e^{-mx}+e^{-m(L-x)}), ~~~
{\tilde F}'_{0\alpha \beta} = \left[ {\begin{array}{cccc}0&0&0&\frac{\xi B_0}{2} \\
0 &0& 0&0\\
0& 0&0&0 
\\ 
-\frac{\xi B_0}{2}&0&0&0\\ \end{array}}\right] \,(e^{-mx}+e^{-m(L-x)}).
 \eqf  
 We find that vacuum birefringence effects occur at fourth order in $\xi$ again. The fourth order electromagnetic field is given by
 \bb
\bm A^{(4)}(x)=\bm\alpha[1-(1+iqx)e^{-iqx}]e^{i\omega x}+\frac{i}{2}\kappa B_0^2\omega x\bm{\tilde\beta}\left(2e^{-mL}+f(2mx)+e^{-2mL}f(-2mx)\right)e^{-iqx}e^{i\omega x}, \label{solA41}
\ee
where $\bm{\tilde\beta}=\hat y\alpha_y+\hat z\frac{7}{4}\alpha_z$ and $f(\rho)=(1-e^{-\rho})/\rho$. This leads to an ellipticity and rotation given by
\bb
|\varepsilon|=\frac{3}{16}\xi^4\kappa B_0^2\omega x|\cos(qx)|\left(2e^{-mL}+f(2mx)+e^{-2mL}f(-2mx)\right) \label{ellip2}
\ee
and
\bb
|\delta\theta|=\frac{3}{16}\xi^4\kappa B_0^2\omega x|\sin(qx)|\left(2e^{-mL}+f(2mx)+e^{-2mL}f(-2mx)\right), \label{rot2}
\ee
respectively.

\end{document}